\documentclass[prd,aps,preprint,showpacs,nofootinbib,eqsecnum,superscriptaddress]{revtex4-1}

\usepackage{latexsym}
\usepackage{hyperref}
\usepackage{epstopdf}
\usepackage{amsfonts,amssymb,amsmath,amsthm,amsxtra,mathtools,euscript,mathrsfs}
\usepackage{dsfont}
\usepackage{yhmath}
\usepackage{arcs}
\usepackage{slashed}
\usepackage{cancel}
\usepackage{epsf}
\usepackage{epsfig}

\usepackage{bm}
\usepackage{graphicx,color,xcolor,subfigure}
\usepackage[font=small,labelfont=bf]{caption}
\usepackage{dcolumn}
\usepackage{array}
\usepackage{tabularx}
\usepackage{enumerate}
\usepackage{hhline}

\def\e{\eta}

\def\s{\sigma}
\def\k{\kappa}

\def\vt{\vartheta}
\def\bvt{\bar{\vartheta}}
\def\veps{\varepsilon}

\def\k{\kappa}
\def\k5{\kappa_5}

\newcommand{\beq}{\begin{equation}}
\newcommand{\eeq}{\end{equation}}
\newcommand{\beqa}{\begin{eqnarray}}
\newcommand{\eeqa}{\end{eqnarray}}
\newcommand{\barr}{\begin{array}}
\newcommand{\earr}{\end{array}}
\newcommand{\beqaa}{\begin{eqnarray*}}
\newcommand{\eeqaa}{\end{eqnarray*}}

\newcommand{\ie}{\textit{i}.\textit{e}.}

\begin{document}
\title{Teleparallel Gravity in Five Dimensional Theories}

\author{Chao-Qiang Geng}
\email[Electronic address: ]{geng@phys.nthu.edu.tw}
\affiliation{
Chongqing University
of Posts \& Telecommunications, Chongqing, 400065, China}
\affiliation{Department of Physics,
National Tsing Hua University, Hsinchu 300, Taiwan}
\affiliation{Physics Division,
National Center for Theoretical Sciences, Hsinchu 300, Taiwan}
\author{Ling-Wei Luo}
\email[Electronic address: ]{d9622508@oz.nthu.edu.tw}
\affiliation{Department of Physics,
National Tsing Hua University, Hsinchu 300, Taiwan}
\author{Huan-Hsin Tseng}
\email[Electronic address: ]{d943335@oz.nthu.edu.tw}
\affiliation{Department of Physics,
National Tsing Hua University, Hsinchu 300, Taiwan}

\begin{abstract}
We study teleparallel gravity in five-dimensional spacetime with particular discussions on Kaluza-Klein (KK) and
braneworld theories. We directly perform the dimensional reduction by differential forms. In the braneworld theory, 
the teleparallel gravity formalism in the Friedmann-Lema\^{i}tre-Robertson-Walker  cosmology is equivalent
to GR due to  the same Friedmann equation, whereas in the KK case the reduction of our formulation does not 
recover the effect as GR of 4-dimensional spacetime.
\end{abstract}

\date{\today}

\maketitle

\begin{section}{Introduction}

Extra dimension theory was first studied by Kaluza \cite{Kaluza}
and Klein \cite{Klein}, the so-called KK theory, in order to unify
electromagnetism and gravity by gauge theory. Electromagnetic fields
come from the extra 5th-dimension which is compactified in some
small scale. It is usually used to explain the hierarchy problems
with the effective Planck scale in 4-dimension by the dimensional
reduction.
The \emph{large extra dimension} was proposed by Arkani-Hamed,
Dimopoulos and Dvali \cite{ADD}, referred to the \emph{braneworld} theory.
The theory with a brane as the solitonic solution for a
physical object is inspired from supergravity as well as superstring
theory~\cite{Horava:1995qa}.
The ordinary matter fields are
localized on the \emph{brane} embedded into a spacetime of a higher
dimension called \emph{bulk}. Randall and Sundrum \cite{RS} gave two
braneworld models based on particular \emph{non-factorizable} metrics,
named {\em RS-I} and {\em II} models,
leading to a \emph{warp
extra dimension} between two 3-branes to solve the hierarchy
problem and an compactification to generate 4-dimensional gravity, respectively.

On the other hand, \emph{teleparallel gravity}, which an alternative gravity theory other than GR,
 was first considered by Einstein~\cite{Einstein:ap} in terms of  \emph{absolutely parallelism}. 
The Lagrangian of \emph{teleparallel equivalent to
general relativity} (TEGR)
is referred to as the torsion scalar
$T$. Recently, several types of gravity theories with $T$, such as the teleparallel
dark energy~\cite{Geng11} and $f(T)$~\cite{Ferraro} models, have been used
to  explain the acceleration of the universe.

Extra dimension theories in teleparallel gravity have been
explored in the literature~\cite{deAndrade:1999vq, Barbosa:2002mg,
Fiorini:2013hva, Bamba:2013fta, Nozari:2012qi}. In this article, we
first set up a general geometrical scenario of TEGR and then investigate
the general behavior of the bulk as well as the projected effect on the brane. 
In the calculations, we keep our geometric construction as
general as possible so that it is easy to compare our results with those in~\cite{deAndrade:1999vq, Barbosa:2002mg,
Fiorini:2013hva, Bamba:2013fta, Nozari:2012qi}.
 In particular, we concentrate on
  the KK  (without vector
fields) and the braneworld theories.
The application of  the Friedmann-Lema\^{i}tre-Robertson-Walker
(FLRW) cosmology in the braneworld scenario is also discussed.

\end{section}

\begin{section}{Teleparallel Gravity in Five-Dimensional Spacetime}

We formulate  teleparallel gravity in high dimensions. Our
geometrical settings are given as follows. Let $(M,g)$ be a
4-dimensional spacetime (hypersurface) isometrically embedded into a
5-dimensional spacetime (the bulk) $(V,\bar{g})$ by $f:M \to V$. We
use the convention $dx^M=(dx^{\mu},dx^5)$ for the coordinate dual
basis of $V$ with  capital Latin letters $M, N= 0, 1, 2, 3, 5$ and
Greek letters $\mu,\nu,\ldots=0, 1, 2, 3$ as the coordinate indices
of $M$, and middle Latin letters $i, j, k,\ldots= 1, 2, 3$ as the
spatial indices of $M$. An orthonormal frame (tetrad)
$\bvt^A=(\bvt^a,\bvt^5)$ for $V$ is indexed by capital Latin letters
$A, B, C,\ldots= 0,1,2,3,5$ with Latin letters $a, b, c, \ldots =
0,1,2,3$ for tetrads on $M$, while the middle Latin letters $i, j,
k,\ldots= 1, 2, 3$ share with spatial coordinate
indices\footnote{These spatial indices should cause no confusion as
it can be easily read off from the context.}.

For a tetrad $(e_0,\ldots, e_3)$ on $M$, it can be  naturally
extended  as a tetrad $(\bar{e}_0,\ldots,\bar{e}_3,\bar{e}_5)$ on
$V$, \ie $\bar{e}_a:=f_*(e_a)$, where $\bar{e}_5$ is the unit
normal vector field to $M$. The corresponding coframes are
$(\vt^0,\ldots,\vt^3)$ for $M$ and $(\bvt^0,\ldots,\bvt^3,\bvt^5)$
for $V$ with $f^*(\bvt^a) = \vt^a$. We shall identify $M$ with
$\bar{M}:=f(M)\subset V$, and $\vt^a \in T^*M$ with $\bvt^a \in
T^*\bar{M}$..., etc.,  interchangeably. Quantities with bars, e.g,
$\bar{e}_A$, represent objects viewed in $V$. The metric signature
is fixed as $(-,+,+,+,\veps)$ and the sign of the 5th-dimension is
denoted by $\veps:= \bar{g}(\bar{e}_5,\bar{e}_5)= \pm
1$.\footnote{ Note that $1/\veps = \veps = \pm 1$ is used in the 
calculation.}

In the 4-dimensional teleparallel theory, one uses a tetrad $\vt^a$ to
formulate the gravitational theory. The metric $g$ of $M$ is given by
\begin{equation}
ds^2 = g_{\mu\nu}\,dx^{\mu} \otimes dx^{\nu} = \e_{ab}\,\vt^a
\otimes \vt^b\,,
\end{equation}
where $\e_{ab}$ is the Minkowski metric. Occasionally, one writes
the tetrad  in terms of the local coordinate, such as
$e_a=e^{\mu}_a\,\partial_{\mu}$ and $\vt^a = e^{a}_{\mu}\,dx^{\mu}$.
With a given tetrad $e_a$ on $M$, we can define the Weitzenb\"{o}ck
connection by
\begin{equation}\label{E:Weitzenbock}
\nabla^W_{e_a} e_b \equiv 0, \qquad \mbox{(for all $a$, $b$)}
\end{equation}
and the Weitzenb\"{o}ck connection 1-form $\omega^b_a$
on $M$. It is easy to observe that the Weitzenb\"{o}ck connection
(\ref{E:Weitzenbock}) yields a vanishing curvature since
\begin{equation}\label{E:curvature vanish}
R^d{}_{cab} \, e_d = \nabla^W_{e_a}  \nabla^W_{e_b} e_c -
\nabla^W_{e_b}  \nabla^W_{e_a}  e_c - \nabla^W_{[e_a,e_b]}  e_c
\equiv 0\,.
\end{equation}
However, the connection takes torsion to manifest
properties of  spacetime and  the gravitational effects. The
torsion 2-form on $M$ is given by $T^a = \nabla^W \vt^a =
\frac{1}{2}T^a{}_{bc} \, \vt^b \wedge \vt^c$ with the torsion
components $T^a{}_{bc} = e_b(e^a_{\mu})e^{\mu}_c -
e_c(e^a_{\mu})e^{\mu}_b$. Consequently, the torsion scalar on $M$
is given by
\begin{subequations}\label{E:4D Torsion scalar}
\begin{eqnarray}
T &=& \frac{1}{4}\,T_{abc} \, T^{abc} +
      \frac{1}{2}\,T_{abc} \,T^{cba} -
      \,T^{b}{}_{ba} \, T^{c}{}_{c}{}^{a}\label{E:torsion scalar 1}\\
  &=& \frac{1}{4}\,T_{\mu\nu\sigma} \, T^{\mu\nu\sigma} +
      \frac{1}{2}\,T_{\mu\nu\sigma} \,T^{\sigma\nu\mu} -
      \,T^{\nu}{}_{\nu\mu} \, T^{\s}{}_{\s}{}^{\mu}\,.\label{E:torsion scalar 2}
\end{eqnarray}
\end{subequations}

In the following discussions, we adopt differential forms to reduce
large tensor calculations.
Based on
differential forms, the torsion scalar (\ref{E:torsion scalar 1})
can be rewritten as a 4-form
\cite{Gronwald:1997bx,Obukhov:2002tm}
\begin{equation}\label{E:torsion 4-form}
\mathcal{T}= T_a \wedge \star
\left[ {}^{(1)}T^a - 2 \, {}^{(2)}T^a - \frac{1}{2} \, {}^{(3)}T^a) \right]
:=  -\frac{1}{2}\, T_a\wedge H^a\,,
\end{equation}
where
\begin{eqnarray}
{}^{(1)}T^a &:=& T^a - {}^{(2)}T^a - {}^{(3)}T^a\,,
 \nonumber \\
{}^{(2)}T^a &:=& \frac{1}{3}\vt^a \wedge i_{e_b}(T^b)\,,
  \nonumber \\
{}^{(3)}T^a &:=& \frac{1}{3}\,i_{e^a}(\vt_b \wedge T^b)\,,
 \nonumber  \\
H^a &:=& (-2) \star\left[ {}^{(1)}T^a - 2 \, {}^{(2)}T^a - \frac{1}{2} \, {}^{(3)}T^a \right]
\end{eqnarray}
with $\star$  the
Hodge dual operator in $g$ of $M$ and $i_v (\omega)$
the interior product of $v$ with a $k$-form $\omega$. 
Similarly, the 5-form
torsion scalar $\bar{\mathcal{T}} $ of $V$ is  defined as
(\ref{E:torsion 4-form}), namely
\begin{equation}
\label{E:5D Torsion scalar}
\bar{\mathcal{T}} =  \bar{T}_A \wedge \bar{\star}
\left[ {}^{(1)}\bar{T}^A - 2 \, {}^{(2)}\bar{T}^A
- \frac{1}{2} \, {}^{(3)}\bar{T}^A \right]\,,
\end{equation}
where $\bar{\star}$ is the Hodge dual operator in $(V,\bar{g})$ and
$\bar{T}^A:= \nabla^W \bvt^A = \bar{d}\bvt^A + \bar{\omega}^A_B
\wedge \bvt^B = \bar{d} \bvt^A$ is the torsion 2-form on $V$, in
which the two kinds of differentials $d:\Omega^k(M) \to
\Omega^{k+1}(M)$ and $\bar{d}:\Omega^k(V) \to \Omega^{k+1}(V)$
should be carefully distinguished, along with the requirement
$\bar{d}|_{M}=d$. The gravitational action on $V$ is given by
\begin{equation}
{}^{(5)}S= \int - \frac{\bar{\mathcal{T}}}{2\kappa_5} =
\int - \frac{\bar{T}}{2\kappa_{5}}\,dvol^5\,,
\end{equation}
where $\kappa_{5}=8\,\pi\,G^{(5)}$ represents the 5-dimensional gravitational
coupling, $\bar{T}$ stands for the torsion scalar of $V$, and
$dvol^5= {}^{(5)}e \, d^5x =\det (e_{M}^A) \, d^5x $
is the volume form of $V$.

Since the hypersurface $M$ is at least an immersion of $V$, there
always exists a coordinate system such that we can write the 5D
metric $\bar{g}$ as the form
\begin{equation}
\bar{g}_{MN} =
\begin{pmatrix}
    g_{\mu\nu}(x^{\mu},y)  &  0  \\
    0  &  \veps \phi^2(x^{\mu},y) \,,
\end{pmatrix}
\end{equation}
where $y=x^5$. Within such a coordinate, we have a preferred frame for
$V$ with
\begin{equation}\label{E:special frame}
\bar{e}_A= \left( e_a, \frac{1}{\phi} \, \frac{\partial}{
\partial y} \right), \quad \bvt^A = \left( \bvt^a, \phi  \, dy \right)\,,
\end{equation}
which will be used in this study unless
particularly specified.

\end{section}

\begin{section}{Effective Gravitational Action on the Hypersurface $M$}

The usual high dimensional theory in GR, such as KK and braneworld
scenarios~\cite{Shiromizu:1999wj,Binetruy:1999ut},
uses the \emph{Gauss-Codazzi equation} to relate the equation of motions
between the hypersurface $(M,g,\nabla)$ and the bulk
$(V,\bar{g},\bar{\nabla})$. The crucial ingredient that guides the
physical connection is through the \emph{extrinsic curvature}, defined by
\begin{equation}\label{Def:extrinsic curvature}
K(X,Y):= - \veps \, \bar{g}\left( \overline{\nabla}_{f_*(X)}
\bar{e}_5, f_*(Y) \right) \qquad \qquad (X,Y\in \mathfrak{X}(M))\,,
\end{equation}
However, in TEGR, since the Weitzenb\"{o}ck connection implies
$K_{ab} = 0$ by (\ref{E:Weitzenbock}) and the vanishing
curvature by (\ref{E:curvature vanish}), the Gauss-Codazzi equation
\begin{equation}\label{E:Gauss-Codazzi}
\bar{R}^a{}_{bcd} = R^a{}_{bcd} - \veps K^a_c \, K_{db} + \veps
K^a_d \, K_{cb}
\end{equation}
is simply a zero identity, so that the hypersurface is
like a flat-paper in an Euclidean space $\mathbb{R}^3$.
In fact, we can construct an \emph{extrinsic torsion}
\cite{Gladush:1998xf,Wu:2011kh} similar to the extrinsic
curvature of GR in (\ref{Def:extrinsic curvature}) to describe the
dynamics of the embedded spacetime in TEGR, which is given by:
\begin{eqnarray}
\label{Def:extrinsic torsion}
B(X,Y) &:=& \veps \bar{g} \left( \bar{T}(f_*(X), f_*(Y)) , \bar{e}_5
\right),\qquad \qquad (X,Y\in \mathfrak{X}(M))\,,
\nonumber\\
       &=& \veps \left[ \bar{g} \left( f_*(X) , \overline{\nabla}_{f_*(Y)} \bar{e}_5 \right)
      - \bar{g} \left( f_*(Y) , \overline{\nabla}_{f_*(X)} \bar{e}_5 \right) \right]\,.
\end{eqnarray}
However, since we use tetrads $(\vt^0, \ldots, \vt^3)$ and $(\bvt^0,
\ldots, \bvt^3,\bvt^5)$ with the Weitzenb\"{o}ck connections of $M$
and $V$ in Sec. II, the extrinsic torsion (\ref{Def:extrinsic
torsion}) does not give us more information from the extra-dimension
due to
\begin{equation}
B_{ab} = \veps \bar{g} \left( \bar{e}_a,
\overline{\nabla}_{\bar{e}_b} \bar{e}_5 \right) - \veps \bar{g}
\left( \bar{e}_b, \overline{\nabla}_{\bar{e}_a} \bar{e}_5 \right)
= 0, \qquad \mbox{(for all $a$, $b$)}\,.
\end{equation}

The extra degree of freedom in TEGR actually is contained in a torsion 2-form.
With the general setting above, we can now calculate the projected
effect onto $M$ from $V$. We decompose the torsion $\bar{T}^a$ of
$V$ into normal and parallel components respect to $M$ by
\begin{equation}\label{E:torsion decompose}
\bar{T}^a = T^a + \bar{T}^a{}_{b 5} \, \bvt^b \wedge \vt^5 \,,
\end{equation}
where $T^a=\frac{1}{2} T^a{}_{bc}\, \vt^b \wedge \vt^c$ is the
torsion 2-form on $M$. In the frame (\ref{E:special frame}), the
nonvanishing torsion components of $V$ are $T^{a}{}_{bc}$,
$\bar{T}^{a}{}_{5b} = e_{5}(e^a_{\mu})e^{\mu}_b$ and
$\bar{T}^{5}{}_{b 5}= \frac{1}{\phi} \, e_{b}(\phi)$. In particular,
if we let the ambient space $V$ be a local product of $U \times W$,
where $U \subseteq M$ is an open in $M$ and  $W$
corresponds to the extra spatial dimension. The local product
structure of $V$ allows us to integrate  over the base space $U$ of
$M$,
\begin{equation}\label{E:5D action}
S_{\text{bulk}} = \frac{-1}{2\k5} \int_U \int_W \left(T  +
\frac{1}{2} \left( T_{ab5} \, T^{ab5} + T_{a5b}\, T^{b5a} \right) +
\frac{2}{\phi} e_a(\phi) \, t^a -  t_5 \cdot t^5 \right) \phi dy \,
dvol^4 ,
\end{equation}
 where $T$ is the (induced) 4-dimensional torsion scalar
defined in (\ref{E:4D Torsion scalar}). The equation in~(\ref{E:5D action})
 provides us with the general
effective action for the hypersurface $M$ in TEGR. In the next, we concentrate on
two specific theories of braneworld and
Kaluza-Klein scenarios.

\begin{subsection}{Braneworld Scenario}
In the braneworld scenario, we set the hypersurface $M$ located at
$y=0$ as a brane and specifying $V = M \times \mathbb{R}$ as a
 product manifold.  From~(\ref{E:5D action}),
 the action on the bulk reads
\begin{equation}\label{E:bulk action}
S_{\text{bulk}} = \frac{-1}{2\k5} \int_M \int_{\mathbb{R}}
\left\{\phi T + \phi \left( \frac{1}{2} \left( T_{ab5} \, T^{ab5} +
T_{a5b}\, T^{b5a} \right) + \frac{2}{\phi} e_a(\phi) \, t^a -  t_5
\, t^5 \right)\right\} \, dy \, dvol^4
\end{equation}
The first term of the parentheses in (\ref{E:bulk action}), 
recognized as $\int_M \int_{\mathbb{R}}\phi \, T \, \sqrt{-g} \, dy
\, d^4x$, is the usual TEGR Lagrangian with a nonminimal
coupled  scalar field $\phi$ on the brane localized in the
5th-dimension, which is equivalent to the nonminimal
coupled Hilbert action $\int_M
\int_{\mathbb{R}} \phi \, R \, \sqrt{-g} \, dy \, d^4x$ of
4-dimension in GR. The second term arises from the 5th-dimensional
component.

According to the \emph{induced-matter theory}, the 5th-dimensional
component and the flow along the 5th-dimension of the second term in (\ref{E:bulk action})
can be regarded as the induced-matter 
from \emph{geometry}. 
It is the projected effect due to the extra spatial dimension.
We note that the \emph{mathematically equivalent} formulations between
the induced-matter and braneworld theories have been demonstrated by Ponce de Leon
in~\cite{PonceDeLeon:2001un}. In Sec.~IV, we shall
discuss 5-dimensional field equations and the corresponding
braneworld cosmology.

\end{subsection}

\begin{subsection}{Kaluza-Klein Theory}

In the KK theory, we take the space $V$ locally as $U
\times S^1$ and consider the 4-dimensional effective low-energy
theory to obtain the \emph{KK ansatz} in TEGR, that is
\begin{equation}
\label{E:Cylindrical condition}
e_5(g_{\mu\nu}) =0 \qquad \mbox{or}\qquad \frac{\partial}{\partial
y}g_{\mu\nu} =0
\end{equation}
with only the \emph{massless Fourier mode} \cite{Overduin:1998pn}.
The metric is reduced to
\begin{equation}
\bar{g}_{MN} =
\begin{pmatrix}
    g_{\mu\nu}(x^{\mu})  &  0  \\
    0                    &  \phi^2(x^{\mu})
\end{pmatrix}
\end{equation}
with $\veps=+1$. Due to the \emph{KK ansatz},
we have $T^a{}_{b5} =0$ and $t^5=0$ so that the extra-dimensional integration is trivial,
\ie
$\int_{S^1} \phi(x^{\mu})\,dy=2\pi r\, \phi(x^{\mu})$, 
where $r$ is the radius of the 5th-dimension. As a result,
we obtain
\begin{equation}
\label{E:effective KK action}
S_{\text{KK}} = \frac{-1}{2\kappa_{4}} \int_U \left( \phi\, T
+ 2\,\partial_{\mu}\phi \, t^{\mu} \right) e \, d^4x ,
\end{equation}
where $\kappa_{4}:= \k5 /2\pi r$ is the \emph{effective}
4-dimensional gravitational coupling constant.
We point out  that our result of (\ref{E:effective KK action})
disagrees with that given in~\cite{Bamba:2013fta}.
One can use a simple case with $F(T)=T$ in Eq.~(5) of~\cite{Bamba:2013fta} to see that the resultant equation
differs from ours in~(\ref{E:effective KK action}).

\end{subsection}

\end{section}

\begin{section}{Friemann Equation of Braneworld Scenario in TEGR}

\begin{subsection}{FLRW Brane Universe}

We now study the teleparallel braneworld effect in
cosmology. We assume that the  brane $M$ with $y=0$ gives a
homogeneous and isotropic universe. The bulk metric $\bar{g}$
is maximally symmetric 3-space with spatially flat ($k=0$), given by
\begin{equation}\label{E:bulk metric}
\bar{g}_{MN} = diag \left(-1, a^2(t,y), a^2(t,y), a^2(t,y),
\veps\,\phi^2(t,y)\right)
\end{equation}
by choosing a coframe with $\bvt^0 = dt$, $\bvt^i = a (t,y)\, dx^i$ and
$\bvt^5 = \phi(t,y) \, dy$. Subsequently, the torsion 2-forms are
\begin{equation}
\bar{T}^0= \bar{d} \, \bvt^0 = 0,\qquad \bar{T}^i = \bar{d} \bvt^i =
\frac{\dot{a}}{a} \, \bvt^0 \wedge \bvt^i + \frac{a'}{a\phi} \,
\bvt^5 \wedge \bvt^i, \qquad \bar{T}^5 =  \frac{\dot{\phi}}{\phi} \,
\bvt^0 \wedge \bvt^5\,,
\end{equation}
where the \emph{dot} and  \emph{prime} stand for the partial
derivatives respect to t and y, respectively.
The torsion scalar in (\ref{E:5D Torsion scalar})
reads
\begin{equation}
\bar{\mathcal{T}} = \left[ T + \left(\frac{3-9\,\veps}{\phi^2}\,
\frac{a'^2}{a^2} + 6\, \frac{\dot{a}}{a} \, \frac{\dot{\phi}}{\phi}
\right) \right] dvol^5
\end{equation}
where $\veps=+1$ and $T = 6\dot{a}^2/a^2$ is the usual 4-dimensional
scalar torsion.
\end{subsection}

\begin{subsection}{Equations of Motion}
The gravitational field equations on the bulk can be derived from the
formulation given
in~\cite{Gronwald:1997bx,Obukhov:2002tm}. The equations
of motion on $V$ are 4-forms
\begin{equation}\label{E:4-form EOM}
\bar{D}\bar{H}_A - \bar{E}_A = - 2\,\k5\,{}^{(5)} \bar{\Sigma}_A\,,
\end{equation}
with
\begin{eqnarray}\label{Def:H_A,E_A}
\bar{H}_A &=& (-2) \bar{\star} \left( {}^{(1)}\bar{T}_A
  - 2 \, {}^{(2)}\bar{T}_A
  - \frac{1}{2} \, {}^{(3)}\bar{T}_A \right)\,, \nonumber\\
\bar{E}_A &:=& i_{\bar{e}_A}(\bar{\mathcal{T}})
  + i_{\bar{e}_A}(\bar{T}^B) \wedge \bar{H}_B\,,
  \nonumber\\
\bar{\Sigma}_A &:=& \frac{\delta \bar{L}_{mat}}{\delta \bvt^A}\,,
\end{eqnarray}
where $\bar{\Sigma}_A$ is the canonical energy-momentum 4-form of
matter fields, and   $\bar{H}_A$ can be
simplified as~\cite{Obukhov:2002tm}
\begin{equation}\label{E:H_A}
\bar{H}_A = \left( \bar{g}^{BC}\bar{K}^D_C \right) \wedge
\bar{\star}\left( \bvt_A \wedge \bvt_B \wedge \bvt_D \right)\,,
\end{equation}
with $\bar{K}^D_C := \bar{\omega}^D_C - \tilde{\omega}^D_C$ being the
contortion 1-form. Here, $\tilde{\omega}^D_C$ is the Levi-Civita
connection 1-form with respect to the coframe with the unique expression of
$\tilde{\omega}^D_C$, given by
\begin{eqnarray}
\label{E:Levi-Civita connection}
\tilde{\omega}^0_i &=& \,\, \frac{\dot{a}}{a} \, \bvt^i,  \quad
\tilde{\omega}^i_0 = \,\, \tilde{\omega}^0_i ,\quad
\tilde{\omega}^0_5 = \veps \frac{\dot{\phi}}{\phi} \, \bvt^n ,
\quad \tilde{\omega}^5_0 = \veps \, \tilde{\omega}^0_5\,,\nonumber\\
\tilde{\omega}^5_j &=& - \veps \frac{a'}{\phi a} \, \bvt^j, \quad
\tilde{\omega}^j_5 = -\veps \tilde{\omega}^i_j,\quad
\tilde{\omega}^i_j \equiv 0\,.
\end{eqnarray}
 From (\ref{E:H_A}) and (\ref{E:Levi-Civita connection}), we obtain
the equations of  motion of the bulk:
\begin{eqnarray}
\label{E:Friedmann eqn}
\bar{D} \bar{H}_0 - \bar{E}_0
&=& 3 \left[ \left(
    \frac{\dot{a}^2}{a^2} + \frac{\dot{a}}{a} \frac{\dot{\phi}}{\phi}
    \right) - \frac{\veps}{\phi^2} \left( \frac{a''}{a} -
    \frac{a'}{a}\frac{\phi'}{\phi}\right) - \left(\frac{1+\veps}
    {2\phi^2}\right)\frac{a'^2}{a^2} \right] \bar{\star}\bvt_0
\nonumber\\
& & +
    \frac{3\veps}{\phi} \left( \frac{\dot{a}'}{a} - \frac{a'}{a}
    \frac{\dot{\phi}}{\phi}\right) \bar{\star}\bvt_5 = - \k5 \,\bar{\Sigma}_0\,,
\nonumber\\
\bar{D} \bar{H}_5 - \bar{E}_5
&=& \frac{3}{\phi}\left(
    \frac{a'}{a}\frac{\dot{\phi}}{\phi} - \frac{\dot{a}'}{a}\right)
    \bar{\star}\bvt_0 + 3\left[ \left( \frac{\ddot{a}}{a} +
    \frac{2\dot{a}^2}{a^2}\right) - \left(\frac{1+\veps}{2\phi^2}\right)
    \frac{a'^2}{a^2} \right] \bar{\star}\bvt_5 = - \k5 \, \bar{\Sigma}_5\,.
\end{eqnarray}
Note that the first equation in Eq.~(\ref{E:Friedmann eqn})
is the Friedmann equation of the bulk.
If we write $\bar{\Sigma}_A = \bar{T}_A^B \, \bar{\star}
\bvt_B$,
we get that
\begin{equation}\label{E:Friedmann eqn 2}
\left( \frac{\dot{a}^2}{a^2} + \frac{\dot{a}}{a}
\frac{\dot{\phi}}{\phi} \right) - \frac{1}{\phi^2} \left(
\frac{a''}{a} - \frac{a'}{a}\frac{\phi'}{\phi}\right) -
\frac{1}{\phi^2} \frac{a'^2}{a^2} = \frac{\k5}{3} \bar{T}_{00}
\end{equation}
with $\veps = + 1$. Furthermore,
if  matter is a perfect fluid, one can decompose the energy-momentum
tensor into bulk and brane parts as~\cite{Binetruy:1999ut}
\begin{eqnarray}
\bar{T}_A^B(t,y) &=& \left(\bar{T}_A^B\right)_{\text{bulk}}+
\left(\bar{T}_A^B\right)_{\text{brane}}\,,
\nonumber\\
\left(\bar{T}_A^B\right)_{\text{brane}} &=& \frac{\delta(y)}{\phi} \,
diag(-\rho(t),P(t),P(t),P(t),0)\,,
\end{eqnarray}
where $\left(\bar{T}_A^B\right)_{\text{bulk}}$ represents as the
\emph{vacuum energy-momentum tensor} or the \emph{cosmological constant}
$(\Lambda_5/\k5)\e_A^B$ in the bulk, and $\rho(t)$ and $P(t)$ are
the energy density and the pressure of the normal matter localized
on the brane, respectively.


For a discontinuous first derivative of the bulk metric $\bar{g}$
($\in C^1(M)\setminus\bigcup_{k> 1}C^{k}(M)$), 
the Dirac delta function would appears in its second derivative,
so that
the FLRW metric could lead to the equation of the scale factor with the form at $y=0$
\begin{equation}
a''(t,y) = \delta(y) \, [a'](t,0) + \tilde{a}''(t,y)\,,
\end{equation}
where $\tilde{a}''$ denotes the non-distributional part of $a''$
and the definition of the \emph{jump} is
\begin{equation}
[f](0):= \lim_{\delta \to 0^+}  f(\delta) - f(-\delta) \qquad (f: M
\to \mathbb{R})\,,
\end{equation}
which measures the discontinuity of a real-valued function $f$ across the
brane. With the form of the scale factor, (\ref{E:Friedmann eqn 2})
yields the junction condition
\begin{equation}\label{E:Jump of scale factor}
[a'](t,0) = \frac{\k5}{3\veps} \rho \, a_0(t) \, \phi_0(t)
\end{equation}
where $a_0(t):= a(t,0)$ and $\phi_0(t):=\phi(t,0)$ are considered as the
scalar factor and a scalar field on the brane, respectively. Furthermore, if we
impose the so-called $\mathbb{Z}_2$ symmetry ~\cite{Horava:1995qa}
for the scale factor in Eq.~(\ref{E:Jump of scale factor}) 
on the bulk as a real-valued quantity $f$ must be an odd function $f(x) = -f(-x)$
across the brane, we obtain the Friedmann equation on the brane to be
\begin{equation}
\label{E:Brane Friedmann eqn} \frac{\dot{a}_0^2(t)}{a_0^2(t)} +
\frac{\ddot{a}_0(t)}{a_0(t)} = - \frac{\k5^2}{36}\, \rho(t) (\rho(t)
+ 3P(t)) - \frac{k_5}{3\phi_0^2(t)} \,
\left(\bar{T}_{55}\right)_{\text{bulk}}\,,
\end{equation}
which is the same as
the braneworld theory of GR shown in~\cite{Binetruy:1999ut}.
Hence,
we confirm that the cosmological braneworld scenario in TEGR coincides
with that of GR, \ie, there is no distinguish between TEGR and GR in
the braneworld FLRW cosmology, which again justifies the name of TEGR.

The physical consequence of the cosmological brane scenario here
then follows from the discussions in~\cite{Binetruy:1999ut}. In particular, if the extra
5th-dimension is compact, one can check if the solutions of
$a(t,y)$ and $\phi(t,y)$ derived from (\ref{E:Friedmann eqn}) are
well-defined  ones, as given in \cite{Binetruy:1999ut}.

Finally, we remark that the Friedmann equation (\ref{E:Friedmann
eqn}) in the bulk can be identified as $G_{00} = - \k5 \bar{T}_{00}$
and $G_{05}=0$, which are the same as those
in~\cite{Binetruy:1999ut}. This result implies that a
\emph{radiating} contribution of the universe can be generated in
TEGR due to the extra spatial dimension. It can be viewed as a
generic property that there exists a component of \emph{dark
radiation} in the braneworld scenario. We have to mention that there
is no extrinsic curvature in TEGR since the projected effects of the
dark radiation and discontinuity property of the brane come from
torsion itself, which is clearly beyond the expectations of
GR~\cite{PonceDeLeon:2001un,Maartens:2010ar} as already pointed out in~\cite{Nozari:2012qi}.

\end{subsection}
\end{section}

\begin{section}{Conclusions}

We have studied teleparallel gravity in five-dimensional spacetime.
In particular, we have shown that the Kaluza-Klein theory in
teleparallel gravity do not generate a Brans-Dicke type of the
effective 4-dimensional Lagrangian as GR. This result is different
from that given in~\cite{Bamba:2013fta}. We have demonstrated that
the braneworld theory of teleparallel gravity in the FLRW cosmology
provides an equivalent viewpoint as Einstein's general relativity.
The additional radiation of the universe can arise from the extra
dimension, which is a generic feature  in the branworld theory.
In GR, the extrinsic curvature plays an important role
to give the projected effects in the lower dimension, which
indicates the embedding can lead to different dynamics of the
hypersurface. On the other hand, in TEGR the projected effect on the
lower dimensional manifold is determined by the projection of
torsion 2-forms.

\end{section}


\begin{acknowledgments}
We are grateful to Professor Friedrich W. Hehl for the encouragement.
We would like to thank Keisuke Izumi, Yen-Chin Ong and Yi-Peng Wu for useful discussions.
The work was supported in part by National Center for Theoretical Sciences,  National Science
Council (NSC-101-2112-M-007-006-MY3) and
National Tsing-Hua University (102N2725E1),  Taiwan, R.O.C.
\end{acknowledgments}


\end{document}